\documentclass[aip, jmp,amsmath,amssymb,
reprint]{revtex4-1}

\usepackage{bm}
\usepackage{amsmath}
\usepackage{amssymb}
\usepackage{xspace}
\usepackage{array}

\usepackage{graphicx}

\newcolumntype{C}[1]{>{\centering\let\newline\\\arraybackslash\hspace{0pt}}m{#1}}

\begin{document}

\title[First principles water from cluster to bulk]
{Toward a universal water model: \\
First principles simulations from the dimer to the liquid phase 
}
%
\author{V. Babin}
\altaffiliation{Contributed equally to this work}
\author{G.R. Medders}
\altaffiliation{Contributed equally to this work}
\author{F. Paesani}
 \email{fpaesani@ucsd.edu}
\affiliation{ 
Department of Chemistry and Biochemistry, University of California, San Diego
\\ La Jolla, California 92093
}%

\date{\today}

\begin{abstract}
A full-dimensional molecular model of water, HBB2-pol, 
derived entirely from ``first principles'', is introduced 
and employed in computer simulations ranging from the dimer 
to the liquid. HBB2-pol provides excellent agreement with 
the measured second and third virial coefficients and, by 
construction, reproduces the dimer vibration-rotation 
tunneling spectrum. The model also predicts the relative 
energy differences between isomers of small water clusters 
within the accuracy of highly correlated electronic structure 
methods. Importantly, when combined with simulation methods 
that explicitly include zero-point energy and quantum thermal 
motion, HBB2-pol accurately describes both structural and 
dynamical properties of the liquid phase. The predictive power 
of the HBB2-pol quantum simulations opens the door to the 
long-sought molecular-level understanding of water under 
different conditions and in different environments.
\end{abstract}

\maketitle

Given the central role that water plays in nature, it is not surprising 
that many studies have attempted to develop a microscopic picture 
of its unique properties. In particular, since the advent of computer 
simulations, myriad molecular models based on both force fields 
(including different degrees of empiricism) and ab initio approaches 
have been proposed. Force field-based models range from coarse-grained
representations with no atomistic details (e.g, see Ref.~\onlinecite{Molinero_V_2009}), 
to classical 
parameterizations in terms of point charges and rigid bonds (see Ref.~\onlinecite{Vega2011}) 
 for a recent review), to yet more sophisticated models that account 
 for molecular flexibility (e.g., Ref.~\onlinecite{Dang_LX_87}), electronic polarization (e.g., Ref.~\onlinecite{Fanourgakis2008a}), 
 and charge transfer\cite{Lee_AJ_2011}. On the other hand, due to the associated 
 computational cost, purely ab initio models so far have been limited 
 to the use of density functional theory (DFT) (e.g., Ref.~\onlinecite{Laasonen_K_1993}). However, 
 despite much recent progress, none of the existing models is capable 
 of correctly describing the properties of water from isolated molecules 
 and small clusters in the gas phase up to the liquid and solid phases. 
 Due to the broad spectrum of (often conflicting) predictions derived from 
 these models, the microscopic behavior of water under different conditions 
 and in different environments remains the subject of continuing debate 
 \cite{Wernet_P_2004,Clark_GNI_2010,Pieniazek_PA_2011,Nihonyanagi_S_2011,Kumar_P_2008,Limmer_DT_2011}.
 
Independently of the number of molecules, the ``first principles'' modeling of 
 water within the Born-Oppenheimer approximation builds upon two 
 components: a faithful representation of the electronic potential energy 
 surface (PES) and the proper treatment of the nuclear motion at a 
 quantum-mechanical level. In principle, the multidimensional PES can be 
 accurately approximated at the coupled cluster level of theory including 
 single, double and perturbative triple excitations, CCSD(T)\cite{Raghavachari1989}, which 
 currently represents the gold standard in quantum chemistry. For the 
 nuclear degrees of freedom, quantum dynamics methods based on basis 
 set expansions of the nuclear wavefunction are well suited to the study of 
 small complexes \cite{Meyer_HD_1990}, while simulation approaches based on the 
 path-integral formalism allow the fully quantum-mechanical modeling of 
 water in condensed phases\cite{Paesani_F_2009}. In the end, these two components must 
 be combined in an efficient computational scheme that enables the 
 calculation of statistically converged quantities. 
 
 \begin{figure}
 \includegraphics[width=8.75cm]{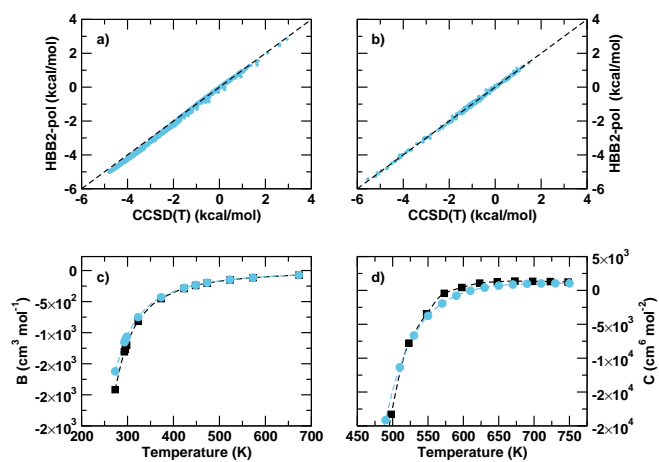}
 \caption{\label{fig:2b3b}2-body (a) and 3-body (b) interactions along with 
     the second (c) and third (d) virial coefficients. Experimental data 
     for second\cite{Harvey2004}  and third\cite{Kell1989} virial coefficients are shown as black squares and 
 the HBB2-pol values as light blue circles.}
 \end{figure}
 
 Unfortunately, the computational cost associated with CCSD(T) makes 
 these calculations prohibitively expensive even for small water clusters. 
 To overcome this computational barrier while still providing an ab initio 
 representation of the molecular interactions, less accurate but more 
 efficient DFT approaches have been widely applied to water simulations. 
 However, the choice of the most appropriate DFT model for water remains 
 the subject of ongoing research \cite{Zhang_C_2011,Lin_I_2012}. Alternatively, the PES for a 
 system of N water molecules can be expressed through the many-body 
 expansion of interaction energies, consisting of 1-body, 2-body, \dots, N-body 
 terms \cite{Hankins1970}. It has been shown that this expansion converges rapidly for 
 water, such that it is sufficient to take into account only the first few 
 \cite{Xantheas1994}. 
 Since the low-order interactions can be accurately calculated using 
 CCSD(T), the many-body expansion effectively enables the representation 
 of the multidimensional PES at the CCSD(T) level of theory \cite{Gora2011}. 
 
 This strategy has been rigorously followed in the development of the CC-pol 
 \cite{Bukowski2007} and WHBB\cite{Wang2011b} models. In CC-pol, which describes the water molecules 
 as rigid, the 2-body term was derived from CCSD(T) data while 
 Hartree-Fock calculations were used to fit the 3-body term. WHBB, which 
 allows for molecular flexibility, was parameterized using CCSD(T) and 
 MP2 reference data for the 2-body and 3-body interaction terms, 
 respectively. Both models include higher-body interactions through point 
 polarizable dipoles. CC-pol accurately reproduce the vibration-rotation 
 tunneling (VRT) spectrum of the water dimer and has been used in 
 classical molecular dynamics (MD) simulations of liquid water \cite{Bukowski2007}. 
 However, these simulations explicitly neglect nuclear quantum effects such 
 as zero-point energy and quantum thermal motion, which have been 
 shown to be important \cite{Paesani_F_2009}. A new version of CC-pol with flexible 
 monomers has recently been applied to study the water dimer \cite{Leforestier2012}. WHBB 
also reproduces the VRT spectrum of the dimer with high accuracy\cite{Leforestier_C_2012}. 
 Due to the computational cost associated with the high dimensionality of 
 the polynomials used to represent the 3-body interactions, quantum 
 simulations with the WHBB model so far have been limited to small clusters \cite{Wang2012b}. 
 
Here, we report on structural and dynamical properties of water calculated 
 with the newly developed full-dimensional model, HBB2-pol, derived 
 entirely from first principles. Similarly to CC-pol and WHBB, HBB2-pol is 
 built upon the many-body expansion of the molecular interactions. The 
 1-body term associated with intramolecular distortion is described by the 
 spectroscopically accurate PES developed by Partridge and Schwenke 
 \cite{Partridge1997a}. The 2-body interaction at short range is represented by the HBB2 
 potential \cite{Shank2009}, which smoothly transitions between 5.5 /AA  and 7.5 /AA into the 
 sum of electrostatic and dispersion interactions, reproducing the correct 
 asymptotic behavior. The induction contributions to non-pairwise additive 
 interactions are taken into account using Thole-type point polarizable 
 dipoles on all atomic sites. In addition, an explicit 3-body component is 
 introduced to account for short-range exchange-repulsion and charge 
 transfer, which have been shown to make a significant contribution to the 
 3-body interaction \cite{Chen1996,Mas2003}. 
 
   \begin{figure}
 \includegraphics[width=8.75cm]{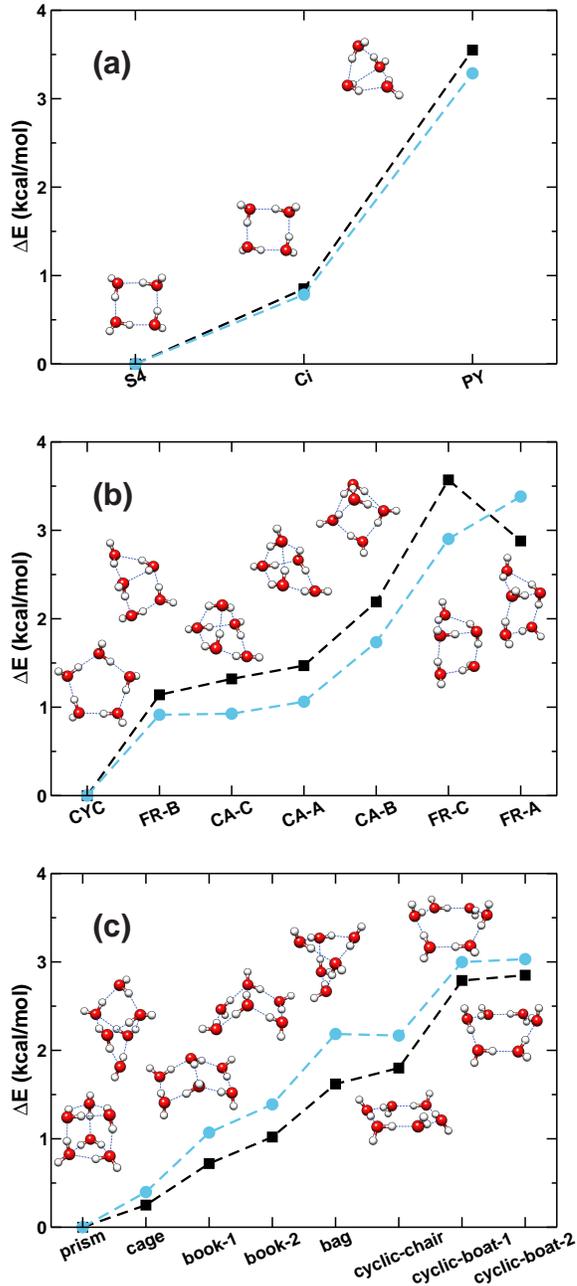}
 \caption{\label{fig:clusters}Relative energies of the low-lying isomers of 
 the water tetramer (a), pentamer (b), and hexamer (c) computed with 
 the HBB2-pol model (light blue circles) and state-of-the-art quantum 
 chemistry methods (black squares) \cite{Temelso_B_2011,Bates2009a}.}
 \end{figure}
 
The 3-body interaction of HBB2-pol is unique in that it is currently the first 
 that has been fitted to CCSD(T) data. Importantly, the inclusion of induction 
 interactions in the 3-body term enables the use of lower-degree 
 polynomials than previously reported \cite{Wang2011b}, resulting in a sizeable decrease 
 in the computational cost associated with the many-body expansion. 
 Specifically, the explicit 3-body component takes the form of a third-degree 
 polynomial in the exponential of the interatomic distances, which is 
 invariant with respect to the permutations of equivalent atoms. This 
 short-range correction was found to be indispensable to overcome the 
 limitations of current polarizable force fields (e.g., Ref.~\onlinecite{Burnham2008}), which attempt 
 to describe all non-pairwise additive interactions through point polarizable 
 dipoles. These findings suggest that 
 the neglect of accurate short-range 3-body interactions invariably leads to 
 an incorrect description of the liquid structure.

The ability of HBB2-pol to accurately reproduce the molecular interactions 
 of water is established by comparison to high-level electronic structure 
 calculations and experimental measurements reported in the literature. 
 Figures~\ref{fig:2b3b}a and \ref{fig:2b3b}b show the correlation plots 
 of the HBB2-pol and 
 CCSD(T) 2-body and 3-body interaction energies, respectively. For this 
 analysis approximately 1400 dimer and 500 trimer configurations were 
 extracted from classical MD simulations of the hexamer, ice Ih, and liquid 
 water. CCSD(T) energies were calculated with the aug-cc-pVTZ basis set 
 \cite{Dunning1989} and corrected for the basis set superposition error through the 
 counterpoise method \cite{Boys1970}. Nearly perfect agreement is found over the 
 entire range of energies, indicating that HBB2-pol accurately reproduces 
 the energetics of both favorably interacting and distorted dimers and 
 trimers. 
 
An additional measure of the overall accuracy of the 2-body and 3-body 
 interactions is provided by the second and third virial coefficients, which 
 directly probe dimer and trimer interactions, respectively. The HBB2-pol 
 results are shown in Figures~\ref{fig:2b3b}c and \ref{fig:2b3b}d along 
 with the corresponding 
 experimental data \cite{Harvey2004,Kell1989}. In the calculation of the second virial 
 coefficient, nuclear quantum effects were explicitly included through the 
 path-integral formalism (see Supporting Information for details). Since 
 experimental data for the third virial coefficient are only available at 
 relatively high temperatures where quantum effects are less important, 
 these calculations were carried out at the classical level with each water 
 monomer held fixed at the ground-state vibrationally-averaged geometry. 
 For both virial coefficients, the calculated values are in close agreement 
 with the corresponding experimental data over the entire ranges of 
 temperature, providing further support of the accuracy of the 2-body and 
 3-body interactions of HBB2-pol.
 
The energetics of small water clusters with more than three molecules 
 allows for a quantitative assessment of the ability of HBB2-pol to correctly 
 describe N-body interactions with N $>$ 3 and represents a stringent test for 
 the overall many-body expansion. For this purpose, the relative energies of 
 the low-lying isomers of the water tetramer, pentamer, and hexamer are 
 compared in Figure~\ref{fig:clusters} with the corresponding ab initio values reported in 
 the literature \cite{Temelso_B_2011,Bates2009a}. Water clusters in this size range are among the 
 largest ones for which highly-correlated electronic structure calculations 
 are still feasible. In all cases, HBB2-pol predicts the correct energy 
 ordering of the isomers, with the energy differences being within the 
 intrinsic accuracy of the ab initio methods \cite{Raghavachari1989}. 
 
The comparisons shown in Figures \ref{fig:2b3b} and \ref{fig:clusters} demonstrate that the degree of 
 accuracy of HBB2-pol is comparable to that of state-of-the-art quantum 
 chemistry methods. While these highly-correlated calculations are 
 restricted to small clusters, quantum simulations with HBB2-pol enable a 
 microscopically detailed characterization of water in condensed phases. 
 To this end, HBB2-pol was used in fully quantum molecular dynamics 
 simulations of liquid water at ambient conditions (T = 298.15~K and 
 $\rho$ = 0.997 g cm$^{-3}$). Specifically, path-integral molecular 
 dynamics (PIMD) and centroid molecular dynamics (CMD) simulations 
 were performed to determine structural and dynamical properties of the 
 liquid phase, respectively \cite{Paesani_F_2009}. Both PIMD and CMD are based on 
 Feynman's formulation of statistical mechanics in terms of path-integrals 
 and have been shown to accurately describe nuclear quantum effects in 
 condensed phases of water \cite{Paesani_F_2009}. All simulations were carried out with a 
 system consisting of 256 molecules in a periodic cubic box.
 
 \begin{figure}
 \centering
 \includegraphics[width=9.75cm]{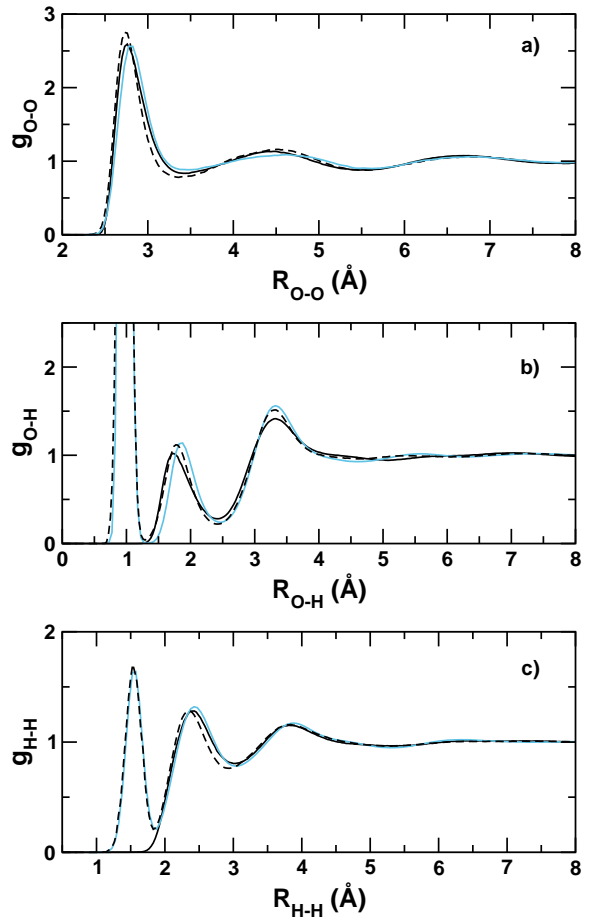}
 \caption{\label{fig:rdf}Oxygen-oxygen, O-O (a), oxygen-hydrogen, 
 O-H (b), hydrogen-hydrogen, H-H (c), radial distribution functions. 
 Experimental data from Refs. \onlinecite{Soper2000,Soper2008} are shown as dashed and 
 solid black lines, respectively. The quantum RDFs calculated with 
 the HBB2-pol model are shown as light blue lines.}
 \end{figure}
 
The oxygen-oxygen (O-O), oxygen-hydrogen (O-H), and 
 hydrogen-hydrogen (H-H) radial distribution functions (RDFs) calculated 
 from an 80 ps PIMD simulation are compared in Figure~\ref{fig:rdf} 
 with two sets of 
 experimental data \cite{Soper2000,Soper2008}. The PIMD simulations correctly predict a lower 
 first peak in the O-O RDF, as determined by the most recent experiments. 
 This feature has been proven difficult to reproduce by current force 
 field-based and ab initio models. Some differences between the PIMD 
 results and the experimental data exist for the second peak of the O-H 
 RDF. However, both position and shape of this peak describing the spatial 
 correlation between O and H atoms directly involved in hydrogen bonds 
 are difficult to be experimentally determined as demonstrated by the 
 appreciable differences between the two sets of experimental data. The 
 diffusion coefficient (D) and orientational relaxation time ($\tau_2$) calculated at 
 the quantum-mechanical level by averaging over six CMD trajectories of 
 10 ps each are listed in Table~\ref{table}. For both quantities, the CMD results are in 
 quantitative agreement with the corresponding experimental values, 
 providing further evidence of the accuracy of the HBB2-pol model. The 
 average structure of the first hydration shell was determined by labeling the 
 water molecules according to the number of donating hydrogen bonds as 
 non-donor, single-donor, and double-donor. Using the geometric criterion 
 proposed in Ref.~\onlinecite{Wernet_P_2004}, the PIMD simulations predict that $\sim$57\% of the 
 molecules are double donors and $\sim$38\% are single donors. Each molecule 
 is involved, on average, in $\sim$3 hydrogen bonds.
 
 \begin{table}
 \caption{\label{table}Comparison between the experimental and calculated diffusion coefficient (D) and orientational relaxation time ($\tau_2$) of liquid water at ambient conditions.}
 \begin{tabular}{C{2.7cm}C{2.2cm}C{2.2cm}}
\cline{1-3}\noalign{\smallskip}
 \multicolumn{1}{C{2.7cm}}{}
 &  \multicolumn{1}{C{2.2cm}}{Experiment}
  &  \multicolumn{1}{C{2.2cm}}{Simulation}\\
\noalign{\smallskip}\cline{1-3}\noalign{\smallskip}
 \hspace{0.5cm}  D ($\text{\AA}^2\text{ps}^{-1}$) & 0.23\footnotemark[1] & 0.23 $\pm$ 0.5 \\
\noalign{\smallskip}
\hspace{0.5cm} $\tau_2$ (ps) & 2.5\footnotemark[2] & 2.5 $\pm$ 0.2 \\
\noalign{\smallskip}\cline{1-3}\noalign{\smallskip}
\end{tabular}
\footnotetext[1]{Ref.~\onlinecite{Krynicki_K_78}}
\footnotetext[2]{Ref.~\onlinecite{Rezus_YL_2005}}
\end{table}
 
In summary, the full-dimensional HBB2-pol model, based entirely on first 
 principles, has been introduced and employed in calculations of the water 
 properties from the dimer to the liquid phase. Being derived from 
 state-of-the-art quantum chemistry calculations, HBB2-pol represents a 
 major step toward the long-sought ``universal model'' capable of describing 
 the behavior of water under different conditions and in different 
 environments \cite{Keutsch2001}. Future simulation studies with the HBB2-pol model will 
 allow the resolution of current controversies regarding structural, 
 thermodynamic, and dynamical properties of bulk, interfacial, and 
 supercooled water.
\end{document}